\documentclass[10pt,conference]{IEEEtran}
\usepackage[T1]{fontenc}
\usepackage{cite}
\usepackage{booktabs}
\usepackage{array}
\usepackage{tabularx}
\usepackage{amsmath}
\usepackage{graphicx}
\usepackage{xurl}
\usepackage[hidelinks]{hyperref}
\newcolumntype{Y}{>{\raggedright\arraybackslash}X}
\title{Test-Driven Approaches to Software Engineering with Large Language Models:\\A Survey of Phases, Tasks, and Agent Skills}
\author{
\IEEEauthorblockN{Yunhao Liang\textsuperscript{1,2}, Chengguang Gan\textsuperscript{3}, Ruixuan Ying\textsuperscript{4},
Hanjun Wei\textsuperscript{2}, Zhe Cui\textsuperscript{1,2}, Shiwen Ni\textsuperscript{5}}
\IEEEauthorblockA{\textsuperscript{1}Chengdu Institute of Computer Applications, Chinese Academy of Sciences, China\\
\textsuperscript{2}University of Chinese Academy of Sciences, China\\
\textsuperscript{3}Independent Researcher, Japan\\
\textsuperscript{4}Institute of Multidisciplinary Research for Advanced Materials (IMRAM), Tohoku University, Japan\\
\textsuperscript{5}Artificial Intelligence Research Institute, Shenzhen University of Advanced Technology, China}
}
\hypersetup{
  pdftitle={Test-Driven Approaches to Software Engineering with Large Language Models: A Survey of Phases, Tasks, and Agent Skills},
  pdfauthor={Yunhao Liang, Chengguang Gan, Ruixuan Ying, Hanjun Wei, Zhe Cui, Shiwen Ni},
  pdfsubject={A scoping survey of test-driven approaches to LLM software engineering}
}
\begin{document}
\maketitle
\begin{abstract}
Tests increasingly participate in the decisions made by large language models and software engineering agents. They specify intended behavior, guide program construction and repair, select candidates, constrain transformations, and provide execution evidence for software analysis. These uses draw on test-driven development, yet differ substantially in test order, oracle availability, editable artifacts, and the role of execution. We present a structured scoping survey organized around the question of what decision a test changes. The review integrates 87 research and supporting records, with method- or protocol-level extraction for 83 records, alongside a separate collection of five practice resources. We distinguish the Red--Green--Refactor cycle from test-conditioned generation, execution-guided refinement, test-mediated analysis, and evaluation-only testing. We then compare code generation, repair, translation, refactoring, clone detection, code search, localization, training-data construction, and formal-specification validation. A dedicated analysis examines how agent workflows and reusable skills encode testing procedures and how their effects are evaluated. Across these tasks, the evidence supports treating test availability, test validity, feedback use, and evaluation independence as separate properties. Test passing alone does not establish behavioral equivalence, effective feedback, or process adherence; aggregate improvements can also conceal different outcomes across models, tasks, and denominators. We synthesize these distinctions into a mechanism taxonomy, a cross-task comparison, and a protocol-sensitive evidence analysis, and identify research directions in oracle validation, causal evaluation, long-horizon maintenance, and reusable test-driven agent capabilities.
\end{abstract}
\section{Introduction}
\label{sec:introduction}

\begin{figure*}[t]
\centering
\includegraphics[width=\textwidth]{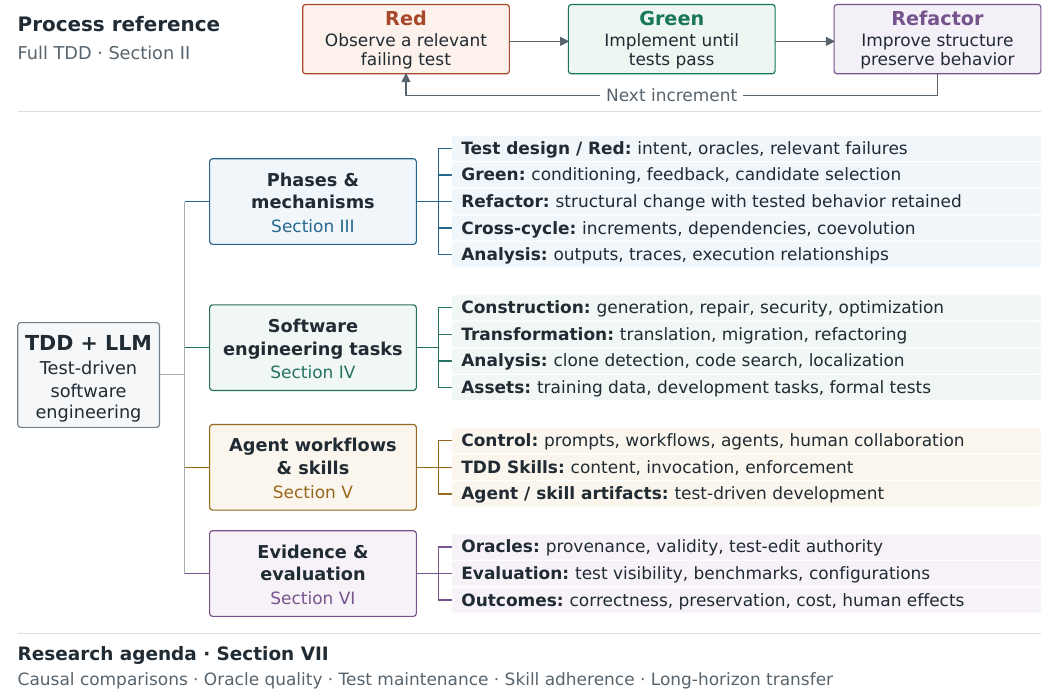}
\caption{Survey taxonomy and reading map. Four complementary dimensions organize the synthesis: TDD phases and test-driven mechanisms (Section III), software engineering tasks (Section IV), agent workflows and skills (Section V), and evaluation evidence (Section VI). Tree branches denote topic membership; individual studies can span multiple dimensions. The process reference above shows the full Red--Green--Refactor cycle defined in Section II. Section VII brings these dimensions together in the research agenda.}
\label{fig:overview}
\end{figure*}

A software test can influence an LLM-generated program before the program exists, during its revision, or only after development has finished. These placements have different consequences. A visible input--output example can resolve an ambiguous requirement. A failing assertion can identify a behavior that a candidate implementation must change. A regression test can reject a transformation that damages existing functionality. When tests are hidden and run only by an evaluator, they instead determine what the reported score measures. Research on test-conditioned generation, generated-test selection, self-debugging, and stronger evaluation suites illustrates these distinct roles \cite{mathews2024testdriven,chen2022codet,chen2023teaching,liu2023is}. Figure~\ref{fig:overview} organizes the survey around phases and mechanisms, software engineering tasks, agent workflows and skills, and evaluation evidence, with the full TDD cycle as a process reference.

Test-driven development (TDD) offers a process-oriented way to connect requirements, implementation, and behavioral preservation. In its Red--Green--Refactor cycle, a developer establishes a relevant failing test, implements enough behavior to satisfy it, and improves the design while retaining passing tests \cite{beck2003test}. LLM-based work applies parts of this reasoning at several scales. LLM4TDD organizes incremental generation around tests; TENET prepares executable test infrastructure for repository-level generation; class-level TDD introduces dependency-aware method scheduling; and TDD-Agent allows tests and implementations to evolve during reasoning \cite{piya2023llm4tdd,hu2025tenet,liang2026scaling,yu2026tddagent}. Their shared use of tests does not make their execution order or oracle assumptions identical.

The connection extends beyond generating and repairing code. Translation can use source execution as a behavioral reference, as illustrated by UniTrans and TransAgent \cite{yang2024exploring,yuan2024transagent}. HyClone uses dynamic behavior to support semantic clone classification, whereas IssueExec uses test traces to locate code relevant to an issue \cite{liang2025hyclone,liu2026issueexec}. In these analysis tasks, the immediate output may be a label or a ranked location rather than a program revision. A survey restricted to whether a system writes tests before code would miss this broader role of executable evidence.

Reusable agent skills add another dimension. A skill may instruct an agent to establish a failure, make a small edit, rerun tests, and inspect regressions. The presence of those instructions, however, is a different observation from their execution or their contribution to task success. SWE-Skills-Bench evaluates the marginal utility of injected skill documents, while work on repository context files investigates how persistent guidance changes agent behavior \cite{han2026sweskillsbench,gloaguen2026evaluating}. These studies motivate separating the contents of a procedure from the mechanism that executes it.

Existing surveys provide substantial foundations. Software-testing and test-generation reviews organize testing activities, generation techniques, and integration strategies \cite{wang2023software,tasarsu2026test}. Program-repair surveys classify repair mechanisms and test-feedback pipelines \cite{zhang2024a,yang2025a}. The Vibe Coding survey discusses development modes, and skill surveys organize representation, acquisition, retrieval, evaluation, and evolution \cite{ge2025a,zhou2026a,ding2026agent}. A critical survey of self-correction further emphasizes the source and reliability of feedback \cite{kamoi2024when}. Table~\ref{tab:surveys} compares their organizing frameworks with the questions pursued here. Our contribution is a common extraction unit across these perspectives: the decision changed by a test, together with the oracle, editable artifact, and evidence of its use. This unit lets us compare a translation reference, a clone-classification observation, and a skill-validation test without assigning them the same development phase.

\begin{table*}[t]
\caption{Positioning against eight related surveys. Entries summarize the inspected organizing frameworks and positive coverage. The final column identifies the complementary question pursued here.}
\label{tab:surveys}
\centering\small
\renewcommand{\arraystretch}{1.1}
\begin{tabularx}{\textwidth}{@{}>{\raggedright\arraybackslash}p{2.7cm}YYY@{}}
\toprule
Related survey & Organizing framework & Treatment of tests or feedback & Complementary question in this survey \\
\midrule
Software testing \cite{wang2023software} & Testing activities and LLM techniques. & Test generation, bug analysis, repair and testing workflows. & Which engineering decision does a test change across tasks? \\
Test generation \cite{tasarsu2026test} & Generation pipeline, data, metrics and integration. & Quality and workflow integration of generated tests. & Who supplies inputs, determines expected outputs, and may revise tests? \\
Program repair SLR \cite{zhang2024a} & Models, repair scenarios, inputs and evaluation. & Tests as repair inputs, feedback and correctness checks. & How do reproduction, patching and non-repair uses of execution differ? \\
Software repair taxonomy \cite{yang2025a} & Fine-tuning, prompting, procedural and agentic approaches. & Explicit test-feedback pipelines and benchmark protocol profiles. & Which phase evidence and oracle assumptions transfer beyond repair? \\
Vibe coding \cite{ge2025a} & Development modes and human--agent interaction. & An explicit test-driven model and iterative feedback. & Which Red/Green/Refactor activities are described and actually measured? \\
Agent skills taxonomy \cite{zhou2026a} & Representation, acquisition, retrieval and evolution. & Skill validation, feedback and runtime governance. & How do TDD content, invocation, adherence and enforcement interact? \\
Skill evaluation \cite{ding2026agent} & Evolution methods and benchmark families. & Execution feedback and task/skill evaluation. & How does using a TDD skill differ from testing the skill artifact? \\
Self-correction \cite{kamoi2024when} & Feedback sources and correction assumptions. & External feedback, including execution, versus intrinsic correction. & How do oracle reliability and editable artifacts change the interpretation of feedback? \\
\bottomrule
\end{tabularx}
\end{table*}

Our review makes three contributions. First, it provides an operational taxonomy that connects TDD phases to test-conditioned construction, feedback-guided revision, candidate selection, and test-mediated analysis. Second, it compares tasks through the artifact being changed, the origin of the expected behavior, and the permissions available to the model. Third, it synthesizes evaluation evidence for both software outcomes and agent procedures, including reusable skills. This organization explains why a generated test can be useful for one purpose yet insufficient for another, and why evidence of successful execution must be interpreted together with the underlying protocol.

Section~\ref{sec:method} defines the scope and review procedure. Section~\ref{sec:mechanisms} develops the mechanism taxonomy, and Section~\ref{sec:tasks} compares task applications. Section~\ref{sec:skills} examines workflows and skills. Section~\ref{sec:evidence} synthesizes evaluation evidence, followed by the research agenda and conclusions.

\section{Background, Scope, and Review Method}
\label{sec:method}

\subsection{Operational Meaning of Test-Driven}

We use \emph{test-driven approaches} as an umbrella term for methods in which tests influence an engineering decision. Within that scope, full TDD retains its narrower process meaning: a relevant failure precedes implementation, passing behavior is established, and structural improvement is checked against preserved behavior \cite{beck2003test}. Test-first preparation is evidence about ordering; it does not by itself establish that a relevant failure was observed. Likewise, repeated code refinement does not establish that refactoring occurred. Refactoring requires a structural objective in addition to a behavioral-preservation objective.

Several neighboring practices require separate treatment. Test-conditioned generation places examples or assertions in the prompt. Test-guided repair uses execution feedback after a candidate exists. Test-based selection ranks or rejects candidates without necessarily editing them. Test-mediated analysis uses outputs, coverage, or traces to decide whether programs are related or where a change should occur. Evaluation-only testing assigns scores after the agent has completed its work. The same paper can contain several such configurations; the classification therefore follows the experimental setting rather than the title alone.

Behavior-driven development (BDD) is relevant when behavioral scenarios connect requirements to executable checks. Exploring Behavior-Driven Development for Code Generation distinguishes a natural-language validation variant from a variant that translates scenarios into tests \cite{liang2025exploring}. We retain that difference: model evaluation of a scenario and execution of a corresponding test supply different kinds of evidence. Similarly, compiler output is execution feedback, but a successful compilation alone does not establish the behavioral property represented by a test.

\subsection{Research Questions and Coding Dimensions}

The review addresses five questions:
\begin{enumerate}
\item Which TDD phases and neighboring test-driven mechanisms are operationalized in LLM software engineering?
\item How do tests express expected behavior, and who generates, validates, and modifies them?
\item How do tests change generation, repair, selection, stopping, classification, retrieval, localization, annotation, and data-construction decisions?
\item How are testing procedures implemented as agent workflows or reusable skills, and what evidence establishes their execution and utility?
\item Which conclusions about correctness, behavioral preservation, analysis accuracy, and engineering cost are supported under the reported evaluation protocols?
\end{enumerate}

The extraction schema contains 33 fields covering bibliographic identity, task, artifact granularity, phase, integration mechanism, test role, test source, oracle origin, test mutability, ordering, phase evidence, evaluation isolation, experimental setting, outcomes, human participation, and evidence location. Test source and oracle origin are separate fields. An LLM can generate inputs while a reference implementation supplies their expected outputs. A repository test can also be an inherited specification whose adequacy differs from that of the hidden evaluation suite.

For phase coding, we use test design, Red, Green, Refactor, and cross-cycle activity. Test design and cross-cycle activity are analytical descriptors around the standard cycle. A non-development analysis task may have no direct phase assignment. Unknown extraction values are distinguished from an inapplicable concept and from evidence that a capability is absent.

\subsection{Search and Source Selection}

We conducted an iterative, structured scoping review with a cutoff of September~9, 2026. Search combined public topic queries, named-method searches, author-provided publication leads, and targeted reference tracing. Query families paired TDD, test-first, test-guided, execution feedback, or executable specifications with LLMs and individual software engineering tasks. Task expansion included translation, migration, clone detection, code search, localization, optimization, specification validation, and training-data construction. Search results were followed to inspectable primary sources, including archival preprints, publisher pages, proceedings, and author-hosted papers.

The retained collection contains 87 research and supporting records. A supplemental search on September~10 used the same publication cutoff and added four studies on execution-informed clone representations and skill development or representation \cite{zhao2023understanding,zhang2026skilltolora,dang2026skiller,lin2026museautoskill}. Its 16 query strings, domain filters, and examined-candidate decisions are preserved with the earlier query logs. Retrieval used ranked web results followed by primary-source inspection; it did not produce a complete bibliographic-database export. Consequently, retained-record totals describe this collection and do not estimate database recall. Five practice resources remain separate from the research records.

Eligibility depends on a test-to-decision relationship. Direct methods use tests or executable observations to influence construction, modification, selection, analysis, or learning. Phase-specific studies and evaluation benchmarks are retained when they establish how to interpret those mechanisms, while related surveys and skill studies provide separately coded context. The working collection uses English-readable primary sources. Earlier code models are retained when they instantiate a relevant mechanism, with their model class made explicit. For example, FuzzTuning's CodeBERT/UniXcoder experiments inform execution-based representation learning rather than autonomous-agent TDD \cite{zhao2023understanding}. Historical non-LLM search, unrelated skill benchmarks, and proposal-only material not selected for the operational comparison are excluded or deferred in the screening ledger. The accompanying protocol records source restrictions and unresolved candidates.

Version handling uses stable paper identifiers and linked publication records. An updated title or publication year does not create another study. Conversely, shared abbreviations are insufficient for merging: the two TDAD papers concern different artifacts and mechanisms \cite{alonso2026tdad,rehan2026testdriven}. Preprint metadata is cited as archival metadata when a formal venue has not been established. Records that contain placeholder conference fields are not treated as evidence of publication at that venue.

\subsection{Extraction and Synthesis}

We assign one primary corpus role per record. The collection contains 43 records on code-construction or modification decisions, six on analysis or data-construction decisions, nine providing phase or test-asset support, ten evaluation-support records, 11 skill-related records, and eight surveys. These categories organize the synthesis; they are not counts of complete TDD implementations.

Method- or protocol-focused passages were inspected for 83 records. Three records remain supported at abstract level, and one tool record relies on primary documentation. Downloading a source and reading it are tracked separately. Method-level extraction supports claims about the reported procedure but does not imply independent reproduction or a complete audit of every result. Abstract-level records are used for narrowly stated scope and mechanism descriptions.

For the coverage map, we recode 56 method/protocol-level records with an operational or phase-specific test role. Each receives one primary task and multiple supported roles, preserving secondary tasks in the full extraction matrix. Surveys, skill/context studies without an assigned operational test role, evaluator-only support, and abstract- or documentation-only records are omitted from this count. Verifier-driven revision of skill artifacts is included when explicitly described, separately from production-code phase evidence. Two test-maintenance studies remain in the review but are not counted because the current extraction does not establish a role within the map's seven categories. A paper-level ledger records every inclusion, exclusion, and role assignment; blank cells denote unassigned coverage rather than demonstrated absence.

Synthesis proceeds in three steps. We first identify the test-to-decision relationship, then compare this relationship across tasks, and finally inspect representative empirical contrasts under their original models, datasets, and denominators. The unit of quantitative comparison is an experimental configuration, not a paper-wide headline. We do not pool heterogeneous pass rates into an overall TDD effect. The accompanying artifacts contain the consolidated review protocol, exact preserved query logs, candidate dispositions, extraction matrix, configuration notes, and claim-to-source ledger. Twelve focused source checks revisit phase order, oracle access, skill intervention and evaluation denominators; a separate comparison records the inspected organizing frameworks of eight surveys. These checks strengthen traceability within the existing workflow.

The review was conducted through a single AI-assisted extraction workflow with iterative source checks; it has not undergone independent duplicate screening or coder-agreement measurement. Its retrieval coverage is therefore scoped rather than exhaustive. Author-supplied leads improve recall for a research line but can also concentrate attention; we apply the same mechanism and evidence criteria to those papers as to other records. These properties constrain claims about prevalence and completeness while still permitting a traceable mechanism-level synthesis.

\section{TDD Phases and Test-Driven Mechanisms}
\label{sec:mechanisms}

\subsection{Test Design and Red: Establishing the Behavioral Target}

The first question is how an intended behavior becomes an executable obligation. In interactive generation, users can accept, reject, or refine examples, giving the system information that the original description omitted. The interactive test-driven generation study separates actual user participation from experiments that simulate responses using a reference implementation \cite{fakhoury2024llmbased}. This distinction matters because reference access resolves questions that a user may find ambiguous or costly to answer. BDD-style scenarios offer another representation of intent, while security tests add properties that ordinary functional examples may not expose \cite{liang2025exploring,liang2026security}.

Existing tests reduce the cost of constructing a target but do not determine execution order. TENET uses a test harness around a missing repository implementation, whereas TDD-Agent permits generation and revision of both tests and code within its workflow \cite{hu2025tenet,yu2026tddagent}. The relevant extraction questions are whether tests existed first, whether they were executable in the initial environment, and whether the observed failure concerned the intended missing behavior. These questions distinguish a declared test-first protocol from an executed Red step.

Bug reproduction makes the validity of a failure particularly important. LIBRO generates tests from bug reports and integrates them into the surrounding project before ranking candidate reproductions \cite{kang2022large}. TDD-Bench Verified and SWT-Bench provide task protocols for evaluating issue-oriented tests and their relationship to fixes \cite{ahmed2024tddbench,mundler2024swtbench}. AssertFlip begins with generated passing tests and inverts their assertions to construct a reproduction target \cite{khatib2025assertflip}. Heterogeneous prompting and execution-feedback selection provide further support for issue-test construction \cite{ahmed2026heterogeneous}. Across these approaches, an uncompilable test, an environmental failure, and a failure exposing the reported defect are different outcomes.

Self-generated tests introduce an additional dependency: the model may be wrong about the expected answer. RECODE uses agreement among generated candidates to improve test reliability and couples this with fine-grained execution feedback \cite{liang2025recode}. Agreement is useful selection evidence, yet correlated candidates may share the same misconception. This yields a general extraction rule: record who proposes the input, who determines the expected output, and what validates that expectation. A single label such as ``LLM-generated test'' obscures these separate responsibilities.

\subsection{Green: Conditioning, Feedback, and Selection}

Tests affect implementation through several mechanisms. In test-conditioned generation, examples constrain a model before it emits its candidate. The TGen study explicitly separates generation without tests, generation with public tests, and remediation using failed tests \cite{mathews2024testdriven}. Tests as Prompt studies a setting in which executable tests themselves serve as generation input \cite{cui2025tests}. These configurations change the information available to the model; an improvement does not by itself identify whether the benefit comes from a test-first process or from a more informative specification.

Execution-guided revision begins after a candidate exists. Self-Debugging studies feedback and explanation configurations, while conversational repair repeatedly exposes failures to a model \cite{chen2023teaching,xia2023keep}. Self-Edit makes this relationship explicit through generation, execution on example tests, and a fault-aware editor \cite{zhang2023selfedit}. ContrastRepair instead constructs contrasting test-case pairs to make the relevant behavioral distinction more informative \cite{kong2024contrastrepair}. RePair uses process-based feedback, and INTERVENOR separates a coding role from a teaching role that interprets errors \cite{zhao2024repair,wang2023intervenor}. Their common structure is a feedback channel, but the content and recipient of that channel differ.

Feedback granularity is one source of variation. A Boolean failure says that a candidate violates an obligation. An assertion message can reveal the observed and expected values. A stack trace points to execution context. Basic-block states and aligned source/target traces can narrow where behavior diverged, as in RECODE and TransAgent \cite{liang2025recode,yuan2024transagent}. Richer feedback can also be longer, noisier, or more expensive. Its value must therefore be compared with the action it enables, rather than inferred from the quantity of information returned.

Candidate selection is a separate alternative to editing. CodeT executes generated solutions against generated tests and uses agreement to select candidates \cite{chen2022codet}. Agentless uses validation signals within a larger localization-and-repair pipeline \cite{xia2024agentless}. Search-oriented methods such as LATS and S* place execution into exploration, verification, or discrimination among candidates \cite{zhou2023language,li2025s}. Work on the exploration--exploitation tradeoff shows why repeatedly repairing one candidate and sampling new candidates consume the same finite resource in different ways \cite{tang2024code}. A fair comparison must account for those resources and for whether a method can recover from an initially poor search direction.

Tests can also influence training. CodeRL learns from code-execution outcomes, Repair-R1 incorporates test and repair signals, and TransCoder-ST uses automated tests in unsupervised translation training \cite{le2022coderl,hu2025repairr1,roziere2021leveraging}. PerfRL combines correctness and performance signals in a small-code-model optimization setting and also filters candidates during inference \cite{duan2023perfrl}. Training-time reference information must be distinguished from information exposed to a deployed agent. A model benefiting from reference-based training does not imply that the same oracle is available when it solves a new task.

\subsection{Refactor: Preserving Behavior While Changing Structure}

Refactor introduces an objective that is not captured by getting a new test to pass. A transformation should improve a structural property while retaining the relevant existing behavior. RefactorAssist explicitly connects agentic refinement to refactoring, while SWE-Refactor and CodeTaste combine execution checks with structural criteria \cite{cordeiro2026refactorassist,xu2026swerefactor,thillen2026codetaste}. The latter combination is essential: leaving the program unchanged may preserve behavior but fail to perform the requested transformation.

Behavioral preservation remains conditional on the observations made. Differential fuzzing compares original and transformed programs on additional inputs and exposes discrepancies missed by existing suites \cite{dristi2026a}. These discrepancies refute equivalence for the observed cases; the absence of a discrepancy over a finite run has a weaker meaning. Consequently, a review should describe test passing as evidence of tested behavior, rather than silently replacing it with universal equivalence.

The object of refactoring also matters. Automated Unit Test Refactoring changes test code, and Testing Framework Migration changes the infrastructure in which tests are expressed and executed \cite{gao2024automated,alves2026testing}. These tasks improve verification assets but do not directly establish that a production-code agent performs the Refactor phase after each Green step. Their relevant invariants may concern assertion meaning, setup behavior, or framework semantics rather than application functionality alone.

\subsection{Cross-Cycle Organization and Coevolution}

A test-driven process must choose an increment. Function-level methods can often consider a compact interface, whereas class-level generation must schedule methods whose dependencies affect one another \cite{piya2023llm4tdd,liang2026scaling}. AgentCoder divides programming, test design, and execution into roles. AlphaCodium organizes code generation into a flow, and Reflexion carries verbal feedback across attempts \cite{huang2023agentcoder,ridnik2024code,shinn2023reflexion}. These designs differ in where state is retained and how failed attempts affect subsequent actions.

Agentic repair provides more flexible control. ThinkRepair, RepairAgent, and AutoCodeRover use different combinations of reasoning, tools, and execution within repair workflows \cite{yin2024thinkrepair,bouzenia2024repairagent,zhang2024autocoderover}. TDFlow further studies workflows for test-driven development \cite{han2025tdflow}. Such flexibility increases the importance of observing the actual sequence of actions: access to a test tool does not reveal whether it was used before an edit, after it, or only at the end.

Dynamic cogeneration makes both the reproduction test and the patch part of the evolving state \cite{cheng2026dynamic}. This permits adaptation when an initial test is inadequate, but also changes the meaning of eventual success. A passing result after editing the test can reflect an improved specification or a weakened obligation. The distinction requires tracking what changed in the test, why it changed, and whether independent evidence still supports the original requirement.

Long-horizon benchmarks extend this issue across releases and repeated requirements. SWE-CI, SWE-EVO, and SlopCodeBench study maintenance or iterative evolution in which a sequence of locally plausible changes can accumulate broader degradation \cite{chen2026sweci,le2025sweevo,orlanski2026slopcodebench}. They motivate evaluating a growing set of behavioral obligations, rather than treating each successful episode as independent of previous work.

\subsection{Executable Evidence for Analysis and Data Construction}

Not every test-mediated decision belongs to a development cycle. HyClone uses outputs from shared inputs to inform clone classification, CoSQA+ uses execution to support query--code annotation, and IssueExec uses traces from relevant tests to locate code \cite{liang2025hyclone,gong2024cosqa,liu2026issueexec}. Here the test provides evidence for a judgment. A negative classification is not a Red step, and a completed ranking is not a Green implementation.

A further distinction is whether the observation changes a current judgment or the representation used for future judgments. FuzzTuning uses fuzzed input/output examples to enrich code-model inputs and learning for clone retrieval and classification \cite{zhao2023understanding}. Its execution signal is informative without a missing implementation or a failing requirement test. This places behavior-informed learning alongside direct execution-based analysis, while keeping both separate from the Red--Green--Refactor process.

LLM4CBI generates and validates witness programs whose execution spectra support compiler-bug localization \cite{tu2023isolating}. SWE-Flow instead uses runtime dependency information to construct development tasks and training examples \cite{zhang2025sweflow}. These examples broaden the interpretation of a test beyond an assertion returning pass or fail. The useful evidence can be an output relationship, a coverage set, a trace hierarchy, or a dependency relation. Their shared contribution is to connect language-based reasoning to an observable property of software execution.

\begin{table*}[t]
\caption{Mechanism taxonomy. Examples instantiate the listed relationship; inclusion does not establish a complete TDD cycle.}
\label{tab:mechanisms}
\centering\small
\renewcommand{\arraystretch}{1.12}
\begin{tabularx}{\textwidth}{@{}>{\raggedright\arraybackslash}p{2.35cm}YYY@{}}
\toprule
Mechanism & Test-to-decision relationship & Evidence required & Illustrative records \\
\midrule
Red--Green--Refactor & A relevant failure precedes implementation; passing behavior constrains structural improvement. & Ordered observations of all three activities, including the reason for failure and the refactoring objective. & TDD provides the reference process \cite{beck2003test}; phase coverage is coded separately for each study. \\
Test-conditioned generation & Assertions or examples constrain the initial candidate. & Tests visible before generation; source of expected outputs. & TGen; Tests as Prompt; TENET \cite{mathews2024testdriven,cui2025tests,hu2025tenet}. \\
Execution-guided refinement & A candidate's execution result guides a subsequent revision. & Feedback channel, revised artifact, stopping rule, and budget. & Self-Debugging; RepairAgent; UniTrans \cite{chen2023teaching,bouzenia2024repairagent,yang2024exploring}. \\
Test-based selection & Executions determine which candidate is retained. & Candidate population, selection criterion, and oracle assumptions. & CodeT; interactive code generation \cite{chen2022codet,fakhoury2024llmbased}. \\
Test-mediated analysis & Outputs, coverage, or traces drive a relation, relevance, localization, or annotation decision. & Mapping from execution evidence to the analysis output and its evaluation. & HyClone; CoSQA+; Issue-Exec \cite{liang2025hyclone,gong2024cosqa,liu2026issueexec}. \\
Evaluation-only testing & Tests assign a score after the agent finishes. & Test access boundary; absence of an iterative feedback channel in the evaluated setting. & RepoMod-Bench; stronger functional evaluation \cite{li2026repomodbench,liu2023is}. \\
\bottomrule
\end{tabularx}
\end{table*}

\section{Cross-Task Applications and Comparison}
\label{sec:tasks}

\subsection{Code Generation, Requirements, and Security}

Generation begins with an incomplete behavioral description and produces an implementation. The main uncertainty is which program the description permits. Function-level studies provide controlled settings for examining the contribution of examples, while class and repository settings add dependencies, state, and environmental constraints \cite{mathews2024testdriven,cui2025tests,liang2026scaling,hu2025tenet}. The unit of generation therefore affects both what a test can express and what must already exist for that test to execute.

Interactive generation allows a user to clarify the intended behavior. Automated application generation instead distributes requirements, test construction, and implementation across agent roles \cite{fakhoury2024llmbased,wan2025automatically}. GAI4-TDD provides a tool-oriented integration of generative AI with TDD \cite{cassieri2024generative}. These forms of integration expose different costs: obtaining a trustworthy specification may require user effort, reference access, role coordination, or additional model calls. Treating all tests as a free input removes an important part of the engineering problem.

Behavioral specifications can add requirements that are difficult to infer from a prose prompt alone. BDDCoder distinguishes natural-language scenarios from executable-test validation, and security-specification work distinguishes functional obligations from security obligations \cite{liang2025exploring,liang2026security}. The distinction between a model understanding a test and following its rule is equally important. The paired-intervention study holds inputs fixed while changing expected outputs between two compatible specifications, then checks generated behavior on unseen inputs \cite{liang2026do}. This supplies a direct test of specification use, rather than relying solely on an aggregate benchmark score.

The relevant success criteria consequently include more than visible pass rate. They include behavior on unexposed inputs, joint satisfaction of functional and security requirements, and sensitivity to intended changes in the specification. These outcomes answer different questions. An agent may pass supplied examples without implementing the intended general rule, or implement the general rule while failing because the execution environment is incomplete.

\subsection{Bug Reproduction, Repair, and Performance Optimization}

Repair starts from existing behavior and a request to change part of it. Tests can establish the defect, help locate relevant code, guide a patch, or screen candidate fixes. LIBRO and AssertFlip focus on constructing useful reproductions; SWT-Bench connects issue tests with fix validation \cite{kang2022large,khatib2025assertflip,mundler2024swtbench}. The reproduction itself is an engineering artifact whose executability and relevance determine the feedback available to the repair process.

Conversational and autonomous repair differ in how they act on that feedback. A conversational loop can return failure information after each proposed patch, while agentic systems may interleave code inspection, localization, editing, and execution \cite{xia2023keep,bouzenia2024repairagent,zhang2024autocoderover}. ContrastRepair changes the test evidence supplied to the conversation, whereas Agentless emphasizes a structured pipeline and patch validation \cite{kong2024contrastrepair,xia2024agentless}. Comparing these approaches requires distinguishing improvements in locating the defect from improvements in editing or selecting the patch.

Performance optimization adds a second objective to functional preservation. PerfRL uses execution-based correctness and timing signals to rank optimized code \cite{duan2023perfrl}. A faster candidate that violates required behavior is unacceptable, while an unchanged correct candidate may make no optimization progress. This resembles refactoring in its preservation requirement but differs in its explicit performance objective. The review therefore records the optimized property rather than classifying every semantics-preserving edit as the same task.

The scope of regression checks is another important axis. A patch can satisfy the newly introduced failure while breaking unrelated behavior. Graph-based TDAD attempts to provide relevant regression-test context \cite{alonso2026tdad}. Its evaluation illustrates why severity, incidence, and issue resolution should all be measured: a method can reduce the number of failing tests without reducing the fraction of patches that cause any regression.

\subsection{Translation, Migration, Refactoring, and Maintenance}

Translation has a particularly useful oracle: an executable source program. UniTrans generates test inputs, obtains expected behavior from source execution, and uses tests in translation and repair \cite{yang2024exploring}. TransAgent augments final-output checking with fine-grained execution alignment \cite{yuan2024transagent}. These methods can use the source to expose differences that are difficult to infer from target-language syntax alone. Their assurance is nevertheless relative to the source behavior represented by the chosen inputs.

SpecTra validates multiple specification modalities before using them for translation. Static specifications and descriptions can be checked by regenerating source-language code and executing existing tests; dynamic specifications are checked against source execution \cite{nitin2024spectra}. This provides an intermediate point between raw source prompting and repeated target-code repair. The tests first influence which specification is trusted, and that specification subsequently conditions translation. TransCoder-ST provides a historical training-oriented counterpart through test-based filtering in unsupervised translation \cite{roziere2021leveraging}.

Repository migration introduces dependencies between translation units. DepWareTrans orders batches using dependency information and checks the mixed-language repository after each translated batch \cite{chand2026depwaretrans}. Its setting relies on languages that can coexist during migration. The resulting feedback can identify integration failures that would be invisible when evaluating isolated functions. Test-framework migration presents a different problem because the verification infrastructure itself changes \cite{alves2026testing}.

Refactoring evaluations need both behavioral and structural criteria. SWE-Refactor checks developer-derived transformations, while CodeTaste combines repository tests with checks for requested code-pattern changes \cite{xu2026swerefactor,thillen2026codetaste}. Differential fuzzing offers an additional observation channel when existing suites fail to expose a semantic discrepancy \cite{dristi2026a}. These evaluations address whether a requested transformation occurred and whether the observed behavior was retained; neither criterion subsumes the other.

Maintenance extends these obligations through time. SWE-CI studies continuous-integration settings, SWE-EVO uses release-sized changes, and SlopCodeBench examines long-horizon iterative tasks \cite{chen2026sweci,le2025sweevo,orlanski2026slopcodebench}. In these settings, a test suite is both an accumulated specification and an asset that must itself be maintained. A process that succeeds on a single issue can still incur growing costs through redundant tests, fragile fixtures, or changes that make subsequent tasks harder.

RepoMod-Bench supplies an important evaluation boundary. It uses implementation-agnostic tests through standardized interfaces and hides the test suites from the agents \cite{li2026repomodbench}. The tests measure modernization outcomes; their existence does not establish that they guided the agent. This contrast helps separate the quality of the evaluation from the presence of a test-driven method.

\subsection{Semantic Code Clone Detection}

Clone detection asks whether two implementations represent related or equivalent functionality. HyClone combines an LLM judgment with dynamic comparison using generated inputs \cite{liang2025hyclone}. Rather than requiring the model to infer equivalence solely from code text, execution provides behavioral observations that can confirm or challenge its initial interpretation.

The immediate decision is a code-pair label. Test generation must produce inputs that are meaningful for both programs, and the evidence depends on which behaviors those inputs expose. Divergent outputs can supply a concrete counterexample, whereas agreement on sampled inputs leaves unobserved behavior unresolved. Classification precision and recall therefore capture a different objective from the pass rate of a generated implementation. This task motivates retaining software-analysis methods within the review while assigning them no artificial Red or Green phase.

Execution can also enter clone detection through the learned representation rather than an explicit comparison of two programs' outputs. FuzzTuning concatenates fuzz-generated input/output examples with source code when adapting CodeBERT and UniXcoder \cite{zhao2023understanding}. On POJ104, the reported clone-retrieval MAP@R changes from 84.29 to 92.01 for CodeBERT and from 90.52 to 93.40 for UniXcoder. The protocol splits programming problems across training, validation, and test sets. This differs from HyClone in both intervention and outcome: one learns a behavior-enriched representation for retrieval, while the other uses execution evidence in a pairwise classification procedure. Their scores therefore cannot establish which approach is more effective without a shared task and evaluation protocol.

\subsection{Code Search and Relevance Annotation}

Code search links a natural-language request to one or more relevant implementations. CoSQA+ uses test-driven agents to help construct a multi-choice code-search benchmark \cite{gong2024cosqa}. Its pipeline distinguishes straightforward cases from cases requiring executable validation. Tests help assess whether a candidate satisfies the query, and the pipeline can repair the generated test program before making an annotation decision.

The main test-driven artifact is the relevance label. This is different from an online retrieval system executing every candidate for every query. Accordingly, annotation accuracy, agreement with human verification, and downstream retrieval quality should be evaluated separately. The example also reveals a distinct oracle problem: the test must operationalize the query correctly before it can provide evidence about the candidate code.

\subsection{Issue and Compiler-Bug Localization}

IssueExec connects issue descriptions to existing tests and then uses their execution traces to identify relevant code \cite{liu2026issueexec}. Its motivating setting lacks failing tests: the existing tests pass before the issue is resolved. Passing executions remain useful because they reveal which code implements related behavior and how application-level operations traverse infrastructure.

LLM4CBI uses a different signal. It generates witness-program variants, validates them, and collects passing/failing execution information for compiler-bug isolation \cite{tu2023isolating}. The mutated artifact is the test program; the compiler is the localization target. These two methods demonstrate why ``test-driven localization'' is not one uniform protocol. They differ in whether a failure is available, how tests are obtained, and which execution relationship supports the ranking.

Localization should be evaluated at the reported granularity and separately from downstream repair. A better file ranking may reduce search effort without guaranteeing a correct patch. Conversely, an agent may repair a bug despite an imperfect ranking. Keeping these outcomes separate makes the contribution of executable evidence easier to assess.

\subsection{Training-Data and Development-Task Construction}

SWE-Flow uses test execution to recover runtime dependencies and construct software engineering tasks \cite{zhang2025sweflow}. Existing implementations and tests provide a basis for selecting development units, removing implementations, and creating target changes. The resulting dataset supports model training and evaluation.

This is a data-construction use of test-driven reasoning. The recovered sequence is a synthesized development arrangement, rather than an observation that the original developers followed TDD. Evaluation should therefore examine both the fidelity of the constructed tasks and the performance of models trained on them. The method also differs from rewarding a model during training or returning failed tests during inference: execution information shapes the learning problem itself.

\subsection{Formal-Specification Validation}

Tests can validate artifacts more abstract than executable application code. Work on LLM-generated tests for Alloy specifications examines whether generated cases expose incorrect formalizations of natural-language requirements \cite{cunha2026validating}. Its immediate contribution is a verification asset supporting test-driven modeling.

The relevant questions concern whether a test is syntactically valid, whether it represents the requirement, and whether it distinguishes correct from incorrect specifications. This places formal-specification validation close to test design while preserving the identity of the artifact under examination. Evidence that an LLM generates useful tests does not automatically establish an end-to-end process that repairs formal models.

\subsection{What Transfers Across Tasks?}

Across these applications, the transferable abstraction is a relationship between an executable observation and a decision. What varies is the behavioral authority behind the observation. Generation may rely on user examples, translation on source execution, clone detection on differential behavior, localization on coverage or spectra, and formal modeling on requirements and model checking. The authority also determines the interpretation of failure: it can be a violated requirement, a source/target discrepancy, an irrelevant candidate, or a useful witness.

Table~\ref{tab:tasks} summarizes these differences. A workflow transfers successfully only when it preserves the meaning of the test and the decision it informs. Reusing the same prompt sequence across tasks without adapting the oracle can preserve the appearance of TDD while changing its substantive guarantee.

\begin{figure*}[t]
\centering
\includegraphics[width=\textwidth]{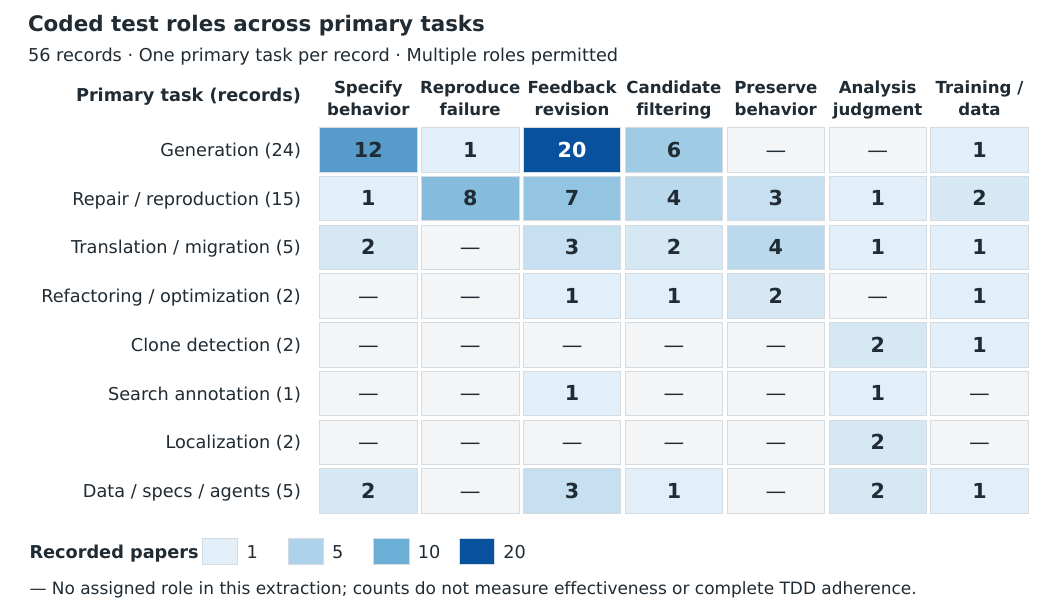}
\caption{Coverage of coded test roles across primary tasks in 56 method/protocol-level records. Each record has one primary task and may contribute to several columns; row totals therefore cannot be recovered by summing cells. Counts include only roles supported by the current extraction, with per-cell paper IDs supplied in the accompanying data. Dashes indicate no assigned role, not absence in the field. Training/data signals are distinguished from execution feedback used during revision. The map describes the retained collection rather than effectiveness or complete TDD adherence.}
\label{fig:coverage}
\end{figure*}

Figure~\ref{fig:coverage} summarizes the resulting coverage by primary task and test role. Of the 56 mapped records, 24 primarily address generation and 15 address repair or reproduction. Feedback-guided revision is prominent in these groups, whereas the analysis rows concentrate on execution-supported judgments. The small analysis groups describe the composition of this collection; they do not establish exhaustive coverage or the size of an unexplored field. Training and data construction form a separate column because an execution-derived learning signal need not be available as feedback during deployment.

\begin{table*}[t]
\caption{Cross-task comparison of the decision changed by tests. Oracle availability and the edited object differ across tasks.}
\label{tab:tasks}
\centering\small
\renewcommand{\arraystretch}{1.12}
\begin{tabularx}{\textwidth}{@{}>{\raggedright\arraybackslash}p{2cm}YYY@{}}
\toprule
Task & Decision and object & Typical oracle or observation & Central distinction \\
\midrule
Code generation & Construct, revise, or select a candidate implementation. & Supplied assertions, generated tests, or human-confirmed outputs. & Test-first access versus observed Red; visible satisfaction versus hidden correctness \cite{piya2023llm4tdd,mathews2024testdriven,fakhoury2024llmbased,liang2026scaling}. \\
Program repair & Reproduce an issue, localize a change, revise a patch, or stop. & Issue text, repository tests, generated reproduction tests. & Test relevance versus patch correctness; product edits versus test edits \cite{ahmed2024tddbench,kang2022large,mundler2024swtbench,bouzenia2024repairagent,han2025tdflow}. \\
Translation and migration & Accept or repair a target program or migrated test. & Source execution, shared inputs, compilation, specifications. & Source-derived behavior versus platform-dependent semantics \cite{roziere2021leveraging,yang2024exploring,yuan2024transagent,nitin2024spectra,alves2026testing}. \\
Refactoring & Accept a structural change or improve test assets. & Regression suite and differential execution. & Passing existing tests versus behavioral preservation; product versus test refactoring \cite{cordeiro2026refactorassist,xu2026swerefactor,dristi2026a,gao2024automated}. \\
Clone detection & Decide whether programs exhibit sufficiently similar behavior. & Shared-input output comparison or fuzz-enriched code representations. & Pairwise judgment versus learned retrieval; neither establishes universal equivalence \cite{liang2025hyclone,zhao2023understanding}. \\
Search and annotation & Validate code--query relevance or construct labels. & Tests that operationalize a query. & Offline annotation versus an online retrieval mechanism \cite{gong2024cosqa}. \\
Localization & Rank issue-related locations or compiler behaviors. & Coverage, execution traces, or compiler test behavior. & Localization evidence versus a subsequent repair loop \cite{liu2026issueexec,tu2023isolating}. \\
Data and specification & Retain training examples, optimize programs, or revise formal artifacts. & Execution filters, rewards, solvers, and behavioral properties. & Retrospective task construction versus observed TDD; solver validity versus requirement validity \cite{zhang2025sweflow,duan2023perfrl,cunha2026validating}. \\
\bottomrule
\end{tabularx}
\end{table*}

\section{Agent Workflows and TDD Skills}
\label{sec:skills}

\subsection{Who Controls the Procedure?}

The integration mechanism determines whether testing is optional, prescribed, or enforced. A prompt can request a sequence of actions. A fixed workflow can schedule those actions externally. An autonomous agent can choose tools and decide when to stop. A multi-agent design can assign test generation and implementation to different roles. Human collaboration adds another source of decisions about requirements and acceptable changes. Application-generation, repair, and workflow studies instantiate different combinations of these choices \cite{wan2025automatically,mock2024generative,huang2023agentcoder,bouzenia2024repairagent,han2025tdflow}.

This distinction affects how a test-driven method should be evaluated. For a fixed orchestrator, the relevant questions include whether the scheduled test is executable and whether the returned feedback changes the next candidate. For an autonomous agent, the evaluation must additionally establish that it selected and used the testing capability. In a multi-agent system, separate role names do not establish independent reasoning: roles may share a model, context, or an incorrect specification.

Agent skills externalize reusable procedural knowledge. The skill-authoring literature treats the skill as an artifact with an interface, supporting resources, and a behavioral evaluation process \cite{destefanis2026authoring}. Surveys distinguish representation, acquisition, retrieval, evaluation, and evolution, which operate at different points in the lifecycle \cite{zhou2026a,ding2026agent}. A TDD skill can therefore be analyzed both as a representation of a development procedure and as a component selected and executed by an agent.

\subsection{Content, Invocation, and Enforcement}

A useful description of a TDD skill should identify the trigger, behavioral target, required ordering, allowed edits, test commands or discovery process, refactoring objective, and stopping condition. These elements establish what the procedure asks the agent to do. They do not alone establish that the environment prevents deviations. A natural-language instruction to observe a failure and an external check that blocks an edit have different operational consequences.

The five practice resources illustrate substantive differences in what a TDD skill prescribes. The Superpowers and Addy Osmani documents include an explicit Red--Green--Refactor sequence \cite{practiceG01,practiceG03}. The inspected Matt Pocock document instead assigns refactoring to a separate review stage and uses a Red--Green loop with one behavioral slice at a time \cite{practiceG02}. The label TDD thus does not uniquely determine phase coverage even among written skill protocols.

Superpowers' skill-authoring guide applies test-driven reasoning to instruction development: observe undesired agent behavior without the skill, write the skill, and assess the changed behavior \cite{practiceG04}. Probity describes operation-level checks that can block actions, placing enforcement outside the skill text \cite{practiceG05}. These are design descriptions, not independently reproduced effectiveness results. The inspected documents are dated snapshots of mutable repository content; their procedural claims should not be generalized to every version.

Repository context files provide relevant comparative evidence. Evaluating AGENTS.md studies conditions with no context file, model-generated context, and developer-provided context, together with traces of testing and exploration \cite{gloaguen2026evaluating}. Its results show why behavior and utility need separate measurement: instructions can increase test execution and tool use without a corresponding improvement in issue resolution. Skill presence, invocation, adherence, and benefit are successive properties to observe, rather than interchangeable descriptions of the same event.

Graph-based TDAD demonstrates a more specific intervention. It combines test-oriented guidance with information about impacted tests and code relationships \cite{alonso2026tdad}. The design question is consequently broader than whether the instruction says ``use TDD'': it also concerns whether the agent receives the test context needed to act on that instruction. Comparing text-only guidance with guidance plus retrieved execution context can isolate this difference.

\subsection{Effectiveness, Cost, and Optimization}

SWE-Skills-Bench reports different outcomes for two TDD-related skills. With Claude Code and Haiku~4.5, the \texttt{tdd-workflow} condition changes pass rate from 21.4\% to 28.6\% on 14 tasks and increases average token cost from 83K to 148K. The \texttt{springboot-tdd} condition changes pass rate from 80.0\% to 70.0\% on ten tasks while reducing tokens from 374K to 236K \cite{han2026sweskillsbench}. These are small, setting-specific contrasts. They establish neither a universal benefit nor a universal cost penalty for TDD skills.

SkillsBench provides broader task coverage, and SkillMOO examines multi-objective optimization of skills \cite{li2026skillsbench,gong2026skillmoo}. These studies support evaluating multiple outcomes but cannot be directly substituted for a TDD-specific treatment comparison. Inside the Skill Market characterizes the activities represented in skill collections rather than testing the effectiveness of every skill it observes \cite{cao2026inside}. Corpus prevalence, benchmark utility, and optimization results therefore belong to different evidence categories.

Skill representation is another intervention variable. Skill-to-LoRA replaces the runtime body of a selected skill with a skill-specific adapter trained from synthetic, skill-conditioned demonstrations \cite{zhang2026skilltolora}. In its Qwen3.6-27B/OpenCode setting, the \texttt{tdd-workflow} row reports one, one, and four solved tasks out of ten for no skill, full skill text, and the adapter, respectively. The ten-task subset differs from the 14-task Haiku~4.5 comparison above, and the runtime cost measure is per-step tokens with offline synthesis and training excluded. This result concerns how a given procedure is represented and activated; the experiment supplies the skill identity and does not establish retrieval accuracy or Red--Green--Refactor adherence.

Guardrails Beat Guidance supplies another relevant distinction: the content of an instruction and the effect of adding instructional context need not coincide \cite{zhang2026guardrails}. Its controlled analysis uses a selected discriminative task subset and rule-content comparisons. Such findings motivate matched controls, but their scope should remain tied to the evaluated selection and agent configuration. A generic rule-file benefit cannot identify which part of a TDD procedure caused an improvement.

\subsection{Developing Agent and Skill Artifacts with Tests}

Testing can also be used to construct the agent artifact itself. Test-Driven AI Agent Definition introduces roles for deriving behavioral tests, revising system prompts, and assessing tests through behavioral mutations \cite{rehan2026testdriven}. Its evaluation separates visible and hidden behavior and includes specification evolution. The object undergoing revision is the system prompt, not the production-code implementation.

This reverses the usual relationship: a test-driven procedure is used to improve the definition of an agent that will later execute tasks. The approach is relevant to skill engineering because both concern reusable behavioral instructions. However, transfer to a specific \texttt{SKILL.md} package requires an evaluation of that artifact's invocation and use. The two TDAD studies in this review accordingly remain distinct: one addresses regression control in code editing, while the other addresses construction of tool-using agent definitions.

Test-driven skill development now includes edits to packaged instructions and helper code. MUSE-Autoskill describes creating a skill package, running its local tests, revising it on failure, and registering it for reuse \cite{lin2026museautoskill}. Its package audit reports a \texttt{tests/} directory in 9\% of generated packages, so the described gate and observed package coverage must remain separate. The generation experiment also distinguishes 35 tasks that yield a skill from the full 51-task population: the reported conditional score is 87.94\%, while the full-population score is 60.35\%. Skills are created from successful runs and re-evaluated on the same tasks. These measurements characterize generation coverage and reuse; new-task generalization requires a different evaluation.

SKILLER optimizes textual skill bundles through verifier-grounded feedback while keeping the executor model fixed \cite{dang2026skiller}. A critic interprets trajectories and verifier diagnostics, and an actor changes instructions or task-local helpers. Its selected \texttt{springboot-tdd} example illustrates successive edits to exploration budgets and artifact consistency. This qualitative case demonstrates what changes in a skill, but does not provide a separate average effect for that skill. Moreover, the optimizer can access a reference trajectory and benchmark feedback. On the skill benchmarks, one instance is used to construct the skill before evaluation across that task's instances; the separately reported held-out GAIA and EarthBench splits answer a different transfer question.

Together, these approaches expose two levels of testing: tests can govern the code-producing agent's behavior, or evaluate the reusable artifact that conditions that behavior. A successful package test, a useful instruction rewrite, and a correctly repaired repository are outcomes at different levels. Reporting the link between them requires both artifact-level validation and task-level evidence.

\section{Evidence, Evaluation, and Validity}
\label{sec:evidence}

\subsection{What Does Each Benchmark Establish?}

Benchmarks observe different parts of the proposed process. HumanEval, MBPP, and their expanded tests primarily assess the behavior of generated functions. EvalPlus demonstrates how strengthening the evaluation suite can expose failures that the original tests miss \cite{liu2023is}. A passing candidate is consequently evidence relative to an oracle and input set. It is not a direct measure of the procedure that produced the candidate.

Issue-level test benchmarks ask a different question. TDD-Bench Verified and SWT-bench evaluate whether a generated test distinguishes an unresolved issue from a repaired version \cite{ahmed2024tddbench,mundler2024swtbench}. They can establish the relevance of a reproduction test without establishing that an agent subsequently repaired the issue. Conversely, a repair benchmark can establish issue resolution without showing that the agent first constructed a failing test. A study must observe both transitions before attributing the result to a complete test-first repair cycle.

Tests as Prompt, CodeTaste, and RepoMod-Bench further illustrate differences in test access \cite{cui2025tests,thillen2026codetaste,li2026repomodbench}. When tests are supplied to the model, performance measures its ability to satisfy an available executable specification. Hidden tests instead assess generalization beyond the checks available during construction. RepoMod-Bench evaluates modification with tests that do not supply an iterative agent feedback channel. Its results therefore inform task difficulty and evaluation design, rather than the effectiveness of an execution-guided loop.

InterCode makes interactive execution an explicit part of the coding environment, allowing evaluations to observe exchanges between an agent and executable feedback \cite{yang2023intercode}. This setting complements terminal correctness scores by exposing the interaction that produced the result.

Long-horizon settings add another dimension. SWE-CI, SWE-EVO, and SlopCodeBench evaluate sequences of changes in which existing behavior and later requirements can interact \cite{chen2026sweci,le2025sweevo,orlanski2026slopcodebench}. Their contribution to this survey is the opportunity to observe accumulated regressions and maintenance behavior. Success on a single isolated change does not determine how test assets, dependencies, or earlier design decisions affect the next change.

\subsection{Comparisons Depend on the Experimental Configuration}

Table~\ref{tab:evidence} retains representative comparisons at the configuration level. The rows illustrate different evidential questions; their metrics are not pooled. A change in pass rate, the fraction of affected tests that fail, and success in following a changed test rule have different denominators and meanings.

Graph-based TDAD provides a particularly instructive contrast \cite{alonso2026tdad}. On its 100-task Phase~1 subset with Qwen3-Coder-30B, the baseline, TDD guidance, and graph-augmented TDD conditions resolve 31\%, 31\%, and 29\% of tasks. Their test-level regression rates are 6.08\%, 9.94\%, and 1.82\%, respectively. The corresponding evaluated test counts differ: 9,245, 8,040, and 8,536. The study's instance-level regression rates do not exhibit the same reduction. The evidence thus supports a reduction in the reported test-level failure proportion under the graph-augmented condition, not a general increase in issue resolution or a corresponding reduction in the fraction of regressing instances.

SpecTra supplies a second contrast at a smaller scale \cite{nitin2024spectra}. In its GPT-4o case study on 24 functions from GNU coreutils \texttt{cat}, the baseline, simultaneous use of all specification modalities, and staged SpecTra procedure produce 10, eight, and 11 accepted first-candidate translations, respectively. These counts describe one project case with a bounded compiler-repair procedure. They suggest that the arrangement of specification information matters, while providing limited grounds for predicting effects across repositories or language pairs.

Security-oriented tests can also have setting-dependent effects. In SecTDD's Qwen2.5-Coder-7B results, providing all tests changes hidden joint correctness from 45.5\% to 41.8\% on SALLM, but from 15.6\% to 40.0\% on CWEval \cite{liang2026security}. The SALLM and CWEval conditions contain 11 and nine tasks, respectively, each repeated with five seeds. The direction of the contrast therefore depends on the benchmark. Joint correctness should remain separate from functional correctness and from a security-only score, especially when repair can improve one dimension while damaging another.

\begin{table*}[t]
\caption{Selected configuration-level contrasts. Values are reported within their original settings; percentages across rows are not comparable effect sizes.}
\label{tab:evidence}
\centering\small
\renewcommand{\arraystretch}{1.12}
\begin{tabularx}{\textwidth}{@{}>{\raggedright\arraybackslash}p{2cm}>{\raggedright\arraybackslash}p{3.35cm}YY@{}}
\toprule
Study & Setting and comparison & Reported outcome & Scope of inference \\
\midrule
SWE-Skills-Bench \cite{han2026sweskillsbench} & Claude Code / Haiku 4.5; \texttt{tdd-workflow}; 14 tasks; no skill $\to$ skill. & Pass rate: 21.4\% $\to$ 28.6\%; tokens: 83K $\to$ 148K. & Small skill-specific sample; increased pass rate accompanies increased cost. \\
SWE-Skills-Bench \cite{han2026sweskillsbench} & Same agent/model; \texttt{springboot-tdd}; 10 tasks; no skill $\to$ skill. & Pass rate: 80\% $\to$ 70\%; tokens: 374K $\to$ 236K. & A distinct task subset; not a repeated estimate of the preceding skill's effect. \\
Skill-to-LoRA \cite{zhang2026skilltolora} & Qwen3.6-27B/OpenCode; \texttt{tdd-workflow}; no skill / full text / adapter. & Solved tasks: 1 / 1 / 4 of 10. & Selected subset; runtime per-step token cost excludes offline synthesis/training. No phase-adherence outcome. \\
Graph-based TDAD \cite{alonso2026tdad} & Qwen3-Coder-30B; 100 tasks; baseline / TDD / graph + TDD. & Resolution: 31 / 31 / 29\%; test regression: 6.08 / 9.94 / 1.82\%. & Evaluated test totals differ; test-level and instance-level regression have different behavior. \\
SpecTra \cite{nitin2024spectra} & GPT-4o; 24 \texttt{cat} functions, C $\to$ Rust; baseline / all modalities / staged method. & Accepted first-candidate translations: 10 / 8 / 11 of 24. & One project case; includes the study's bounded compiler-repair procedure. \\
SecTDD \cite{liang2026security} & Qwen2.5-Coder-7B; SALLM (11 tasks, five seeds); B0 (no tests) $\to$ B3 (all tests). & Hidden joint correctness: 45.5\% $\to$ 41.8\%. & Joint functional/security outcome in this benchmark and configuration. \\
SecTDD \cite{liang2026security} & Same model; CWEval (nine tasks, five seeds); B0 $\to$ B3. & Hidden joint correctness: 15.6\% $\to$ 40.0\%. & Direction differs from SALLM; task/model strata must remain visible. \\
Follow Tests \cite{liang2026do} & Qwen3.6-27B; 20 families, 120 paired realizations; explicit-rule controls versus test-driven switching. & Explicit A/B: 100\% each; paired switch: 60.6\%, interval [43.1, 76.7]. & Different measures diagnose capability versus test use; uncertainty is bootstrapped by family. \\
\bottomrule
\end{tabularx}
\end{table*}

Table~\ref{tab:representatives} applies a common schema to 16 studies selected for task and mechanism diversity. It makes visible two distinctions that aggregate performance tables can obscure: a test's source can differ from the authority behind its expected output, and the ability to revise tests can differ from the ability to revise code. The phase column preserves evaluator-only and configuration-dependent evidence and leaves unestablished properties open. In particular, generated-test selection, code repair, and system-prompt revision should not acquire identical phase labels merely because all three use execution.

\begin{table*}[t]
\caption{Representative studies under a shared extraction schema. Sixteen records were selected for mechanism and task diversity, not as a ranking. Phase codes describe production-code R/G/F evidence; analysis and agent-definition activities retain their own artifact types.}
\label{tab:representatives}
\centering\small
\renewcommand{\arraystretch}{1.08}
\setlength{\tabcolsep}{4pt}
\begin{tabularx}{\textwidth}{@{}>{\raggedright\arraybackslash}p{2.7cm}Y>{\raggedright\arraybackslash}p{2.55cm}>{\raggedright\arraybackslash}p{2.1cm}Y@{}}
\toprule
Study / primary task & Tests and expected-output authority & Test modification & R / G / F & Evaluation boundary \\
\midrule
TGen \cite{mathews2024testdriven} \newline Generation & Public benchmark tests; supplied expected outputs & Supplied tests fixed & U / M / U & Public/private check in the designated configuration \\
TENET \cite{hu2025tenet} \newline Repository generation & Repository tests; repository oracle & Selected subset; edit rights U & M / M / U & Agent subset; harness sees full suite \\
Class-level TDD \cite{liang2026scaling} \newline Class generation & Public class tests; benchmark oracle & Public tests fixed & U / M / NA & Private tests assess generated classes \\
TDD-Agent \cite{yu2026tddagent} \newline Generation & Generated tests; model-inferred outputs & Code and tests mutable & U / M / U & Self-tests during reasoning; final project tests \\
CodeT \cite{chen2022codet} \newline Candidate selection & Generated tests; execution consensus & Separate generated test pool & NA / NA / NA & Development selection and benchmark scoring separated \\
TDD-Bench Verified \cite{ahmed2024tddbench} \newline Bug reproduction & Generated issue tests; buggy/fixed comparison & Test generation; no code repair & E / NA / NA & Gold patch hidden in main condition \\
Dynamic cogeneration \cite{cheng2026dynamic} \newline Repair & Generated reproductions; independent fix/test assessment & Workflow-dependent & C / M / NA & Joint correctness separated from self-consistency \\
TDFlow \cite{han2025tdflow} \newline Repair & Human/generated reproductions; repository regression tests & Reproductions locked after creation & M / M / NA & Human-gold and generated-test conditions differ \\
UniTrans \cite{yang2024exploring} \newline Translation & Generated inputs; source-execution outputs & Test-edit rights U & U / M / NA & Generated development and benchmark tests separated \\
SpecTra \cite{nitin2024spectra} \newline Translation & Existing/generated tests; source-execution validation & Test-edit rights U & U / U / NA & Specification validation uses tests; holdout status varies \\
RefactorAssist \cite{cordeiro2026refactorassist} \newline Refactoring & Project tests; original tested behavior & Code changes; test rights U & NA / M / M & Independent stronger behavior checks U \\
HyClone \cite{liang2025hyclone} \newline Clone detection & Generated inputs; differential program outputs & Input generation; later edits U & NA / NA / NA & Code-pair classification; test independence U \\
CoSQA+ \cite{gong2024cosqa} \newline Search annotation & Query-derived tests; human-checked label subset & Test-program repair allowed & NA / NA / NA & Label checking differs from retrieval evaluation \\
IssueExec \cite{liu2026issueexec} \newline Localization & Passing repository tests; execution traces & Test-edit rights U & NA / NA / NA & Localization output; test independence U \\
Graph-based TDAD \cite{alonso2026tdad} \newline Regression-aware repair & Repository tests; impacted-test map & Code changes; static test map & U / M / NA & F2P/P2P outcomes; configuration-dependent access \\
Agent-definition TDAD \cite{rehan2026testdriven} \newline Agent definition & Behavioral tests; execution assertions & Prompt revised; test rights U & NA / NA / NA & Visible/hidden behavioral tests; mutation validation \\
\bottomrule
\end{tabularx}
\par\smallskip
\begin{minipage}{\textwidth}\footnotesize
M: mechanism described in the method; E: established through evaluator-side comparison; C: configuration-dependent; U: not established by the current extraction; NA: not an activity of the studied production-code process. M is not proof of universal trajectory adherence. Green denotes implementation/revision checked against tests, not candidate selection alone. For SpecTra, validated specifications and bounded compiler repair do not by themselves establish a behavioral test-feedback implementation loop. ``Test rights U'' does not imply permission to change tests. F2P/P2P denote fail-to-pass/pass-to-pass tests.
\end{minipage}
\end{table*}

\subsection{Test Availability, Validity, and Use}

Three questions recur across these evaluations. First, are the relevant tests available to the model? Second, are their expected outcomes trustworthy and discriminative for the intended behavior? Third, does the model use their information when making a decision? Better scores after adding tests can combine effects from additional specification content, additional computation, candidate selection, and feedback-driven revision. A test/no-test comparison alone cannot separate these mechanisms.

Do LLMs Really Follow Tests? studies test use through paired behavioral rules \cite{liang2026do}. Its revised protocol uses 20 task families with six realizations per family, repeated under several checkpoints. Qwen3.6-27B follows an explicit description of each rule at 100\% in the reported control conditions, while its paired test-driven switching rate is 60.6\%, with a family-bootstrap interval of [43.1, 76.7]. This contrast distinguishes competence to implement a rule from reliably inferring and switching to that rule from tests. The family structure and uncertainty also matter: repeated realizations are not independent evidence for hundreds of unrelated task families.

The same separation applies to repair. A failure report can be informative while still being ignored, misinterpreted, or overfit. Self-Repair, Self-Edit, INTERVENOR, and work on LLM self-correction examine different forms of revision and external evidence \cite{olausson2023is,zhang2023selfedit,wang2023intervenor,kamoi2024when}. These studies should be compared through their feedback channels and budget controls. An apparent repair gain may partly reflect another sampling opportunity; conversely, an ineffective loop may reflect a weak oracle rather than a general inability to revise code.

Oracle validation becomes especially important when both programs and tests are generated by related models. Agreement can reflect a shared interpretation error. Reference execution, human confirmation, behavioral mutation, and independently constructed evaluation tests provide different checks on that agreement \cite{fakhoury2024llmbased,chen2022codet,rehan2026testdriven}. None is interchangeable with a larger number of generated assertions: more assertions can repeat the same missing assumption.

\subsection{Behavioral Preservation and Human Outcomes}

Behavioral preservation calls for measurements beyond passing the existing suite. Differential fuzzing of LLM refactoring examines a filtered set of 3,538 refactorings from 4,368 generated candidates and reports non-equivalent transformations missed by existing tests \cite{dristi2026a}. The filtering step defines the evaluated population. Results for compilable, executable candidates should not be silently presented as results for every generated attempt. The conditional fraction of non-equivalent cases missed by tests likewise differs from a failure rate over all transformations.

For human collaboration, correctness, effort, and experience should be reported separately. Interactive code generation includes both a real-user study and larger evaluations with a reference-based user proxy \cite{fakhoury2024llmbased}. The proxy supplies useful controlled answers but does not model all the uncertainty, effort, or misunderstanding of a developer. GAI4TDD and the exploratory TDD interaction studies add tool and human-workflow perspectives \cite{cassieri2024generative,mock2024generative,mock2026vibe}. The latter compare small groups and distinct interaction arrangements; their findings motivate further study without establishing a population-wide productivity effect.

These distinctions suggest a common reporting core: the editable artifacts; test and oracle origins; observed phase transitions; visible and hidden evaluation access; task, attempt, and test denominators; execution and token budgets; unsuccessful or excluded attempts; and the uncertainty attached to the comparison. Studies can then add task-specific outcomes, such as classification quality for clone detection or semantic preservation for translation. Such reporting makes an intervention interpretable without imposing one success metric on every software engineering task.

\subsection{Cross-Study Findings and Their Implications}

\paragraph{Test order, information and feedback are distinct interventions}
TGen separates supplying tests from remediation, CodeT uses execution for selection, and the paired-rule study measures whether a model changes its program when the behavior specified by tests changes \cite{mathews2024testdriven,chen2022codet,liang2026do}. Together they show why a test/no-test score difference cannot identify a complete TDD mechanism: the difference may reflect a clearer target, selection among more candidates, or revision after a failure. Conversely, a method can obtain useful execution evidence without observing Red, as in passing-test localization \cite{liu2026issueexec}. The practical consequence is to evaluate test availability, action order, feedback use and final correctness separately. The evidence supports this decomposition, rather than a universal ranking of test-first and test-after workflows.

\paragraph{Oracle authority determines what can transfer across tasks}
UniTrans can execute the source program to obtain a translation reference, TiCoder can ask a user to resolve intent, and CodeT estimates candidate quality through generated-test agreement \cite{yang2024exploring,fakhoury2024llmbased,chen2022codet}. These mechanisms resolve different uncertainties. A source oracle helps preserve observed behavior but does not supply a desired correction when that behavior is defective. Agreement among model-generated artifacts can improve selection without providing an independent account of the requirement. HyClone and FuzzTuning further show that behavior can support classification or representation learning without proving program equivalence \cite{liang2025hyclone,zhao2023understanding}. Transfer therefore depends on the decision and source of expected behavior, not just access to an execution tool.

\paragraph{Outcome granularity can reverse the interpretation of success}
Graph-based TDAD's lower test-level regression proportion does not coincide with higher issue resolution in its Phase~1 comparison; SecTDD's joint-correctness contrast changes direction across two benchmarks \cite{alonso2026tdad,liang2026security}. These are distinct outcomes and task populations, rather than interchangeable replications of one effect. MUSE-Autoskill supplies a related selection effect: performance conditional on producing a skill has a different denominator from performance across all assigned tasks \cite{lin2026museautoskill}. A review should retain the population and unit of each result before drawing a cross-study conclusion. Reporting task completion, behavioral regressions and cost together would make the engineering tradeoff more informative than any single aggregate score.

\paragraph{Skill utility depends on representation and the evaluation boundary}
SWE-Skills-Bench exposes a written skill, Skill-to-LoRA changes its representation, SKILLER revises its contents using verifier feedback, and MUSE-Autoskill validates and reuses generated packages \cite{han2026sweskillsbench,zhang2026skilltolora,dang2026skiller,lin2026museautoskill}. These studies support treating skills as interventions with a construction history, not merely a present/absent feature. Their evidence also has different access boundaries: skill identity can be supplied, task feedback can guide construction, and a generated skill can be tested on the task that produced it. Repeated runs, cross-agent reuse and held-out task transfer should consequently be reported separately. For TDD, the remaining question is which representations and development procedures reliably induce the required phase transitions on new software tasks.

\section{Research Gaps and Agenda}
\label{sec:agenda}

\subsection{Separate the Effects of Tests, Ordering, and Feedback}

The findings in Section~\ref{sec:evidence} motivate decomposing interventions commonly bundled under TDD. A matched study could hold the model, task, initial candidate, and total budget fixed while varying whether tests are supplied before generation, whether a relevant failure must be observed, whether execution feedback is returned, and whether a structural refactoring objective is present. Repeated sampling with an equivalent budget would help distinguish the contribution of feedback from that of another attempt. Such a design would estimate particular mechanism effects rather than an undifferentiated ``TDD effect.''

Process measurement should accompany outcome measurement. A trajectory can record when a test became available, what failed and why, which artifact changed, and whether the accumulated suite was rerun. Failure due to an unavailable dependency should be distinguished from failure that exposes the target behavior. Refactoring should be identified through its structural objective and subsequent behavioral check. These observations would make it possible to compare prescribed TDD, observed adherence, and successful completion within the same experiment.

\subsection{Develop Oracles That Can Discriminate Competing Interpretations}

The test-design problem is partly an information problem. A suite should distinguish plausible but incompatible interpretations of a requirement. Interactive selection, paired-rule evaluation, and mutation-based validation already provide complementary ways of investigating this property \cite{fakhoury2024llmbased,liang2026do,rehan2026testdriven}. Future work can connect them by measuring the behavioral alternatives eliminated by a test and the human or execution cost of obtaining its expected result.

Oracle construction also needs to reflect task-specific asymmetries. Translation often has an executable source program, repair may have only a defect report and a partial suite, and specification validation may have a solver but an uncertain natural-language intention. A useful method would state which source of authority resolves disagreement. Hybrid designs could use reference execution where available, ask a human about unresolved requirements, and reserve independent checks for evaluation. This would allow test generation to express uncertainty without treating model agreement as semantic ground truth.

\subsection{Treat Test Assets as Maintained Software}

The corpus contains substantial work on generating tests, with smaller bodies of work on test refactoring, migration, and evolving project behavior \cite{tasarsu2026test,gao2024automated,alves2026testing,chen2026sweci,le2025sweevo}. These activities belong to the same lifecycle. As requirements change, an inherited assertion can become obsolete, and an agent must distinguish an invalid test from a valid test that exposes a regression.

Longitudinal evaluations could trace these decisions across successive changes. They should retain both an independent behavioral evaluation and a history of test edits, so that a passing suite can be interpreted alongside changes to its requirements. Outcomes could include retained behavior, defect-revealing strength, test maintenance effort, execution cost, and the time spent resolving uncertain failures. The research question is how test quality and agent behavior evolve together, rather than how many tests can be generated for a fixed snapshot.

\subsection{Connect Construction and Analysis Through Execution Evidence}

Clone detection, code search, localization, and training-data construction show that test-driven reasoning extends beyond modifying an implementation \cite{liang2025hyclone,gong2024cosqa,zhang2025sweflow,liu2026issueexec,tu2023isolating,chand2026depwaretrans}. This creates an opportunity for shared infrastructure: a validated execution observation can help identify similar code, locate affected behavior, choose a repair context, or construct a training example.

The transfer is not automatic. A test sufficient to retrieve likely examples may be too weak to establish clone equivalence. A compiler test useful for localizing an optimization issue may not express a desired source-level requirement. Future evaluations should therefore measure the cost and loss of reliability when an execution observation is reused for another decision. A task-spanning benchmark could expose common artifacts while maintaining distinct correctness criteria for retrieval, classification, localization, and modification.

\subsection{Evaluate Skills as Executed Procedures}

TDD skills invite a layered experimental design. First, validate the written artifact against its intended procedure. Second, test whether the agent retrieves or invokes it on appropriate tasks. Third, inspect adherence to the required order and edit permissions. Finally, evaluate behavior and cost on independently assessed tasks. This separates failures of skill content, selection, execution, and task competence.

Matched comparisons could contrast a short prompt, a reusable skill, and an externally enforced workflow containing the same procedural content. Additional conditions could supply relevant test context or vary the agent's authority to edit tests. The mixed TDD-related results in SWE-Skills-Bench and the graph-based TDAD comparisons make these distinctions empirically relevant \cite{han2026sweskillsbench,alonso2026tdad}. Optimization should consider correctness, regression behavior, and cost together, with evaluation tasks kept separate from those used to revise the skill.

Testing the agent definition itself provides a complementary direction \cite{rehan2026testdriven}. Behavioral tests and mutations can evaluate a prompt or skill before deployment, while task trajectories evaluate its subsequent use. MUSE-Autoskill and SKILLER provide concrete mechanisms for refining reusable skill artifacts, while Skill-to-LoRA changes their runtime representation \cite{lin2026museautoskill,dang2026skiller,zhang2026skilltolora}. A comparison should hold procedural content and executor fixed where possible, separate construction tasks from evaluation tasks, and charge skill creation or training costs to the intended reuse horizon. This would reveal whether a procedure that succeeds in its construction setting retains its benefit on new repositories.

\subsection{Study Collaboration and Transfer Beyond Short Tasks}

Human collaboration is central when tests clarify underspecified requirements. Existing interaction studies establish feasible workflows, but their task scale and participant groups leave open questions about sustained adoption \cite{fakhoury2024llmbased,mock2024generative,mock2026vibe}. Longer studies should examine how developers select tests, challenge incorrect expected outputs, review generated changes, and decide when more validation is warranted. Experience with TDD and with AI tools may affect each activity differently.

Transfer across model families, programming languages, and repository structures should be evaluated separately from within-setting optimization. A skill can depend on a test framework or project convention that is absent in another environment. A translation method can exploit a source oracle that a repair task lacks. Reporting these dependencies would help explain successful transfer and identify which procedural elements remain useful across settings.

\section{Conclusion}
\label{sec:conclusion}

Tests influence LLM software engineering through specification, generation, refinement, selection, behavioral comparison, analysis, and agent control. This survey organized 87 research and supporting records around those decisions, distinguishing the Red--Green--Refactor cycle from neighboring uses of executable evidence. The cross-task comparison shows that test order, oracle origin, edit permissions, and evaluation access determine what a reported method actually establishes.

The empirical evidence is configuration-dependent. Supplying tests does not ensure that a model follows them; lower test-level regression does not necessarily imply higher issue resolution; and the presence of a TDD skill does not establish adherence or utility. Future progress depends on observing these intermediate mechanisms and evaluating them with trustworthy oracles, independent outcomes, and explicit denominators. These distinctions provide a basis for developing and comparing test-driven capabilities across code construction, software analysis, and reusable agent procedures.

\bibliographystyle{IEEEtran}
\bibliography{references}
\end{document}